\documentclass[twocolumn,prb,nobibnotes]{revtex4}
\usepackage{mleftright}
\usepackage{graphicx}
\usepackage{tabularx}
\usepackage{capt-of}
\usepackage{color}
\usepackage{subfigure}
\usepackage{multirow}
\usepackage{array}
\usepackage{amsmath,amssymb,amsfonts,textcomp}
\usepackage{color}
\usepackage{subfigure}
\usepackage{multirow}
\usepackage{array}
\usepackage{physics}
\usepackage{booktabs}
\usepackage{array}
\usepackage{makecell}
\usepackage{graphicx}
\usepackage{dcolumn}
\usepackage{bm}
\begin{document}

\title{Electron correlation in 2D periodic systems}

\author{Oinam Romesh Meitei}
\email{oimeitei@mit.edu}
\affiliation{Department of Chemistry, Massachusetts Institute of Technology, Cambridge, MA 02139}
\author{Troy Van Voorhis}%
 \email{tvan@mit.edu}
\affiliation{Department of Chemistry, Massachusetts Institute of Technology, Cambridge, MA 02139}

\date{\today}

\begin{abstract}
  Given the growing significance of 2D materials in various optoelectronic applications, it is imperative to have simulation tools that can accurately and efficiently describe electron correlation effects in these systems. Here, we show that the recently developed bootstrap embedding (BE) accurately predicts electron correlation energies and structural properties for 2D systems.
Without explicit dependence on the reciprocal space sum ($k$-points) in the correlation calculation, BE can typically recover $\sim$99.5\% of the total electron correlation energy in 2D semi-metal, insulator and semiconductors. We demonstrate that BE can predict lattice constants and bulk moduli for 2D systems with high precision. Furthermore, we highlight the capability of BE to treat electron correlation in twisted bilayer graphene superlattices with large unit cells containing hundreds of carbon atoms.
We find that as the twist angle decreases towards the magic angle, the correlation energy initially decreases in magnitude, followed by a subsequent increase. 
We conclude that BE is a promising electronic structure method for future applications to 2D materials.
\end{abstract}

\maketitle

\section{Introduction}

2D materials research has gained significant momentum in the past few years due to these materials' unique electronic, optical, and mechanical properties. These materials consist of a single layer of atoms or molecules arranged in a 2D lattice structure, which offers several advantages over their bulk counterparts. 
They exhibit exceptional mechanical, optical, and electrical properties, making them attractive for applications in various fields, including electronics\cite{2d1}, optoelectronics\cite{2d2, 2d3, 2d4}, energy storage\cite{2d_energy_storage}, and catalysis\cite{2d_catalysis}.
\cite{2d1, 2d2, 2d3, 2d4}. 
Notably, 2D materials have shown promising potential as topical insulators in electronic devices, as photocatalysts\cite{2d_catalysis, ZHAO202078}, and in organic light-emitting diodes (OLEDs)\cite{2d_display, Zhang2021}.
Recently, the concept of twistronics, in which two or more layers of 2D materials are stacked at specific twist angles to create novel electronic states, has opened up new avenues for the application of 2D materials\cite{bilayer_kaxiras, bilayer_kaxiras2, bilayer4, twistronics}. Such twisted layers of 2D materials have the potential to revolutionize the field of electronics\cite{Hennighausen_2021}, quantum computing\cite{Uri2020, Cao2020}, superconductivity\cite{Stepanov2020, Arora2020}, and photonics\cite{Tang2021, 10.1063/5.0070163}.

Electron correlation effects play a crucial role in determining the electronic and optical properties of 2D materials. Unlike bulk materials, where electrons can move freely in three dimensions, 2D materials confine electrons in a reduced dimensionality, resulting in strong electron-electron interactions and reduced screening of Coulomb interactions. Only a handful of ab initio methods can treat electron correlation effects accurately in 2D systems, namely random phase approximation (RPA)\cite{rpa1, rpa2}, Moller-Plesset perturbation theory (MP2)\cite{bartlettCPL2001, gusJCP2001}, coupled cluster singles and doubles (CCSD)\cite{eomCCJCP2005, gruneisJCTC2011, berkelbachJCTC2017}, variational Monte Carlo (VMC)\cite{PhysRevB.43.12943}, and diffusion Monte Carlo\cite{RevModPhys.73.33}. 
However, these methods are computationally demanding and limited to small unit cells. Periodic quantum embedding methods offer a potential solution to the scalability of these methods by offering computations in sub-system space\cite{PhysRevB.76.045107,PhysRevB.98.085138, https://doi.org/10.1002/qua.25801, doi:10.1063/1.4903828, doi:10.1021/acs.jctc.8b00927,doi:10.1063/5.0084040, doi:10.1063/1.469264, doi:10.1021/acs.jctc.0c00576, doi:10.1021/acs.jctc.6b00651, https://doi.org/10.1002/wcms.1357,PhysRevB.20.5345, INGLESFIELD200189}. Notable examples of periodic quantum embedding methods are the density functional embedding theory\cite{GOVIND1998129, CarterJCP2002}, projection-based wavefunction-in-DFT methods\cite{GoodpasterJCTC2018, doi:10.1021/acs.jctc.9b00571},
density matrix embedding theory (DMET)\cite{GusDMETsolid2014, ChanDMETsolid2020, GagliDMETsolid2020, dmet_multireference}, ab initio DMFT\cite{chanDMETDMFT2020}, regional embedding\cite{regionalembeddingberkelbach}, and local embedding schemes\cite{localembeddinggruneis}.

Alternatively, we have recently proposed periodic bootstrap embedding (BE) to treat electron correlation for periodic systems\cite{be1d}.
The approach allows for a flexible partitioning of the system through the use of fragments with overlapping regions, and it has been shown to achieve convergence in the correlation energy as the fragment size increases. Furthermore, BE improves the embedding description by utilizing matching conditions in the overlapping region of the fragments\cite{beJCTC2020,beJCTC2019, beJPCL2019, beJCP2020, beJCP2016}. 
Numerical tests have demonstrated the high accuracy of the method in computing electron correlation energy in 1D systems and its applicability to systems with large unit cell sizes.
BE has also successfully treated electron correlation in molecules and supra-molecular complexes\cite{beMOLPHYS2017, beJCTC2019, beJCTC2020}.

In this paper, we apply periodic BE to describe electron correlation in various 2D semiconducting materials that are important in diverse fields and applications, including energy storage, catalysis, electronics, optoelectronics, and photonics. 
 Using only the mean-field computation at dense reciprocal space $k$-points, periodic BE provides access to correlation energy at the thermodynamic limit (TDL), which avoids finite-size effects at coarse $k$-points. 
 Using CCSD as the choice of correlated solver, periodic BE-CCSD typically recovers $\sim$99.5\% of the electron correlation energy for the 2D systems tested in this work.
 Periodic BE-CCSD can effectively predict lattice constants and bulk moduli, which are important structural and mechanical properties of 2D systems. We further demonstrate the applicability of periodic BE to large periodic superlattice systems by computing correlation energies of twisted bilayer graphene at a twist angle of 6.01$^\circ$ containing up to 364 carbon atoms in the elementary unit cell. The investigations on the 2D systems in this work are largely still a proof-of-concept, as they involve only computations with minimal basis sets. However, the results suggest the broad potential for periodic BE in electron correlation effects in 2D materials.

\section{Computational Methods}

\subsection{Periodic Bootstrap Embedding}

We begin by capturing the main ideas of periodic BE to treat electron correlation in 2D periodic systems\cite{be1d}. Consider partitioning a system into two distinct parts: a fragment comprising atom-centered localized orbitals and a bath encompassing the rest of the system. Due to the Schmidt decomposition technique, the many-body wavefunction of the entire system can be represented as a tensor product in the Hilbert space, consisting of the many-body fragment and the bath states:

\begin{align}
  \ket{\Psi} = \sum^{N_f}_p \lambda_p \ket{f_p}\otimes\ket{b_p}\label{Eq:SD}
\end{align}

with $N_f$ being the number of fragment states, $\ket{f_p}$ and $\ket{b_p}$ the fragment and bath states, respectively, and $\lambda_p$ characterizing the entanglement between the fragment and the bath. This expression considerably reduces the Hilbert space dimension of the system because the number of bath states cannot exceed the (much smaller) number of fragment states.

As the exact many-body wavefunction is not known a priori, BE employs an approximate bath constructed from the Hartree-Fock (HF) wavefunction instead of the actual many-body bath states to describe the embedding\cite{geraldDMET2013, geraldDMET2012}. When periodic boundary conditions are imposed, HF calculations are typically performed in reciprocal space to take advantage of the translational symmetries inherent in the system. Such computations are carried out on a grid of finite crystal momenta or $k$-point in the reciprocal space first Brillouin zone (FBZ). 
A transformation matrix $T^{k,A}$, composed of the coefficients of the fragment and the bath states, can then be utilized to derive the embedding Hamiltonian with the HF bath:

\begin{align}
\hat{H}^A = \sum^{2N_A}_{pq}h^A_{pq}a^{\dagger}_p a_q + \frac{1}{2}\sum^{2N_A}_{pqrs}V^A_{pqrs}a^{\dagger}_pa^{\dagger}_ra_sa_q\label{Eq:frag_hamiltonian}
\end{align}

with,
\begin{align}
h^A_{pq} = \frac{1}{N_k}\sum_k \sum_{\mu\nu}^N T^{k,A}_{\mu p} F^k_{\mu\nu} T^{k,A}_{\nu q} - V^{HF,A}_{pq}\label{Eq:oneelec}
\end{align}
\begin{align}
V^A_{pqrs} = \frac{1}{N_k}\sum_k \sum_{\mu\nu\lambda\sigma}^N T^{k,A}_{\mu p}T^{k,A}_{\nu q}V^k_{\mu\nu\lambda\sigma}T^{k,A}_{\lambda r}T^{k,A}_{\sigma s}\label{Eq:eri}
\end{align}

where $F^k$ is the Fock matrix, $V^k$ is the two-electron integrals, and $V^{HF,A}$ is the HF potential constructed using the HF density matrix in the embedding basis. 
Indices $\mu, \nu, \lambda, \sigma$ denote local orbitals (LOs) and $p, q, r, s$ denote embedding orbitals. $N_k$ is the total number of $k$-points, and $N_A$ is the number of LOs in fragment A.

The above equations reveal that the fragment embedding Hamiltonian has no explicit dependence on the reciprocal space. As a result, $\hat{H}^A$ can be solved with any non-periodic correlated electronic-structure method. This paper focuses on coupled cluster singles and doubles (CCSD) as the correlated fragment solver. Additionally, since the correlated fragment calculation does not rely on reciprocal space, approaching the thermodynamic limit (TDL) in an embedding calculation is relatively inexpensive; it is only necessary to converge the HF calculation to the TDL, and the correlated embedding calculation will also be converged\cite{be1d}.

\begin{figure}[htb]
\includegraphics[scale=0.5]{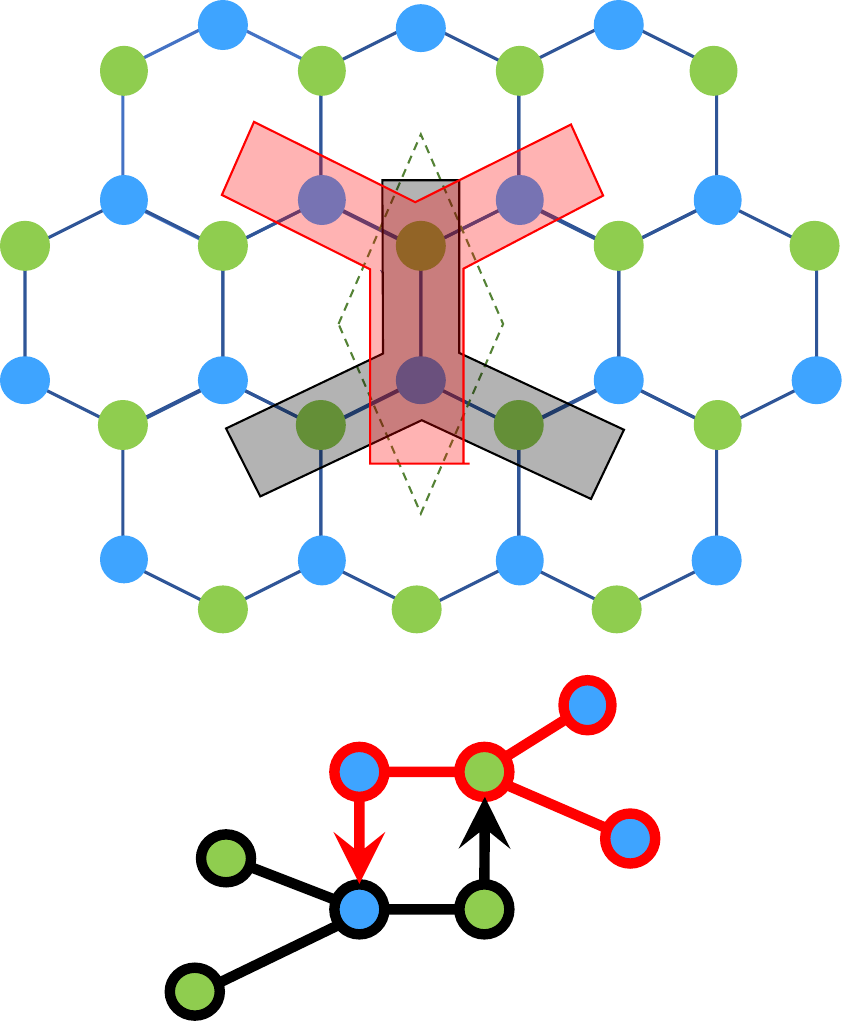}
\caption{Illustration of fragments in periodic BE in a hexagonal lattice system.}\label{Fig:frag-2d}
\end{figure}

The embedding described above can be improved by partitioning the system into fragments that share an overlapped region\cite{beJCP2016}. Figure 1 illustrates such a partitioning scheme for a hexagonal lattice system. The wavefunction at the center of each fragment in the overlapping region is described more accurately than the wavefunction at the edges of the fragment that are adjacent to the HF bath. The reason is that fragment edges more strongly interact with the HF bath compared to the fragment centers. By matching the wave function at the edge of one fragment to the center of another fragment where they overlap, it is then possible to improve the embedding description. The matching of the wavefunction at the overlap region is accomplished in terms of the one-electron reduced density matrix (1-RDM) obtained from solving $\hat{H}^A$ with CCSD method in this work. A global constraint is also imposed on the 1-RDM of each fragment center to conserve the total electron count.

Following our earlier works, we can systematically construct fragments with varying sizes by considering up to the second nearest connected atoms as the fragment edges\cite{beJCTC2020, beJPCL2019}. Henceforth, BE2 denotes fragments that include only the nearest neighboring atoms, and BE3 denotes fragments where the second nearest neighboring atoms are also included. By construction, the number of fragments equals the number of atoms in the unit-cell. For example, there are only two fragments in the honeycomb lattice shown in Figure 1 since the unit-cell contains only two atoms.
When dealing with fragments where all edges extend outside the unit-cell, we construct a superlattice composed of ($2\times2$) unit-cell to accommodate these fragments.

The total BE energy per unit cell is then computed as:

\begin{align}
  E = E_{HF} + \sum^{N_{frag}}_A\sum_{p\in\mathbb{C}_A}\Bigg[\sum_q^{2N_A}F_{pq}^{A,[0]}\Delta P^A_{pq} \nonumber\\
    + \frac{1}{2}\sum^{2N_A}_{qrs}V^A_{pqrs}K^A_{pqrs}\Bigg]\label{Eq:energy}
\end{align}

where $E_{HF}$ is the HF energy, $F^{[0]}$ is the Fock matrix corresponding to the reference HF density in the embedding basis, $\Delta P^A$ is the difference in the correlated 1-RDM and the reference HF density of fragment $A$, and $K^A$ is an approximate two-body cumulant as defined in Ref \citenum{Booth2022}. 
We emphasize the significance of the energy expression in Equation \ref{Eq:energy} for achieving highly accurate correlation energy, in contrast to density-based energy expressions.
In certain cases, employing the energy expression in Equation \ref{Eq:energy} has resulted in a reduction of correlation energy error from approximately 20\% to less than 1\% compared to the density-matrix-based energy used in, for example, refs. \citenum{GusDMETsolid2014}, \citenum{chanDMETDMFT2020}, \citenum{beJCTC2020}, \citenum{iaoGerald2013}, and \citenum{geraldDMET2013}.

\subsection{Twist Averaging}
To establish the accuracy of periodic BE for 2D systems, we compare BE total electron correlation energy per unit cell at the TDL to the full $k$-point CCSD method (hereafter denoted as $k$-CCSD). With the rapidly scaling cost of $k$-CCSD with respect to the system size ($N^6$) and $k$-points ($N^4$), computations were achievable only up to a limited number of $k$-points. The largest $k$-points that were used to compute $k$-CCSD correlation energy in this work was an 18x18 mesh for graphene. Large finite-size errors are well known to occur at small $k$-points. To reduce the finite size error at the coarse $k$-points, the twist-averaging (TA) method was employed in this work to obtain the $k$-CCSD correlation energy\cite{TAmontecarlo, TAShepherd2019, TAShepherd2021, CCAnsatzSolidGruneis2018, TAgruneis2016}.

The TA method involves introducing a set of offsets, also called twist angles, to the unit cell of the crystal lattice. $k$-CCSD computations are then performed for each twist angle, and the resulting correlation energies are averaged over all the twist angles in this expression:

\begin{align}
    E_{corr}^{TA} = \frac{1}{N_s}\sum_{l=1}^{N_s}E_{corr}(k_{s,l})
\end{align}

where $N_s$ is the total number of twist angles, and $E_{corr}(k,l)$ is the $k$-CCSD correlation energy computed with the offset $k_s$.

Averaging the correlation energy over the twist angles enables an improved extrapolation to the thermodynamic limit, with a reduced finite size error for the coarse $k$-points. 
It should be noted that, due to the prohibitive scaling of $k$-CCSD with the number of $k$-points, each of the calculations performed for twist averaging at coarse $k$-points is much cheaper (often by orders of magnitude) than a calculation near the TDL. This effect is partially offset by the need to do several such calculations (typically around 30 in what follows) in order to converge the average. However, overall the improvement in accuracy is substantial. In some cases (most notably graphene) we are unable to extrapolate $k$-CCSD to the TDL without twist averaging.
It should be noted that although the computational cost scales linearly with a factor of $N_s$, a significant number of $k$-CCSD calculations must be performed.

\subsection{Calculation Details}

The accuracy of periodic BE for 2D systems was tested on a set of monolayer 2D systems, namely graphene, hexagonal BN (h-BN), silicon carbide (SiC), triazine-based graphitic carbon nitride (g-C$_3$N$_4$), and molybdenum disulfide (MoS$_2$). Calculations were also performed on a set of twisted bilayer graphene. Further information on the structure of all the systems used in this work is provided in the Supporting Information.

All calculations reported in this work were carried out using a minimal basis set (STO-3G). The $k$-point HF and $k$-CCSF were performed with a Monkhorst-Pack k-point sampling. The BE calculations were performed with an in-house code that uses PySCF to generate the necessary integrals. To obtain the total correlation energy per unit cell at the TDL, a power law expansion was employed to extrapolate the BE and $k$-CCSD energies.

\begin{align}
    E(N_k) = E_{\infty} + \alpha N^{-1}_k
\end{align}

\begin{figure}[htb!]
\includegraphics[scale=0.312]{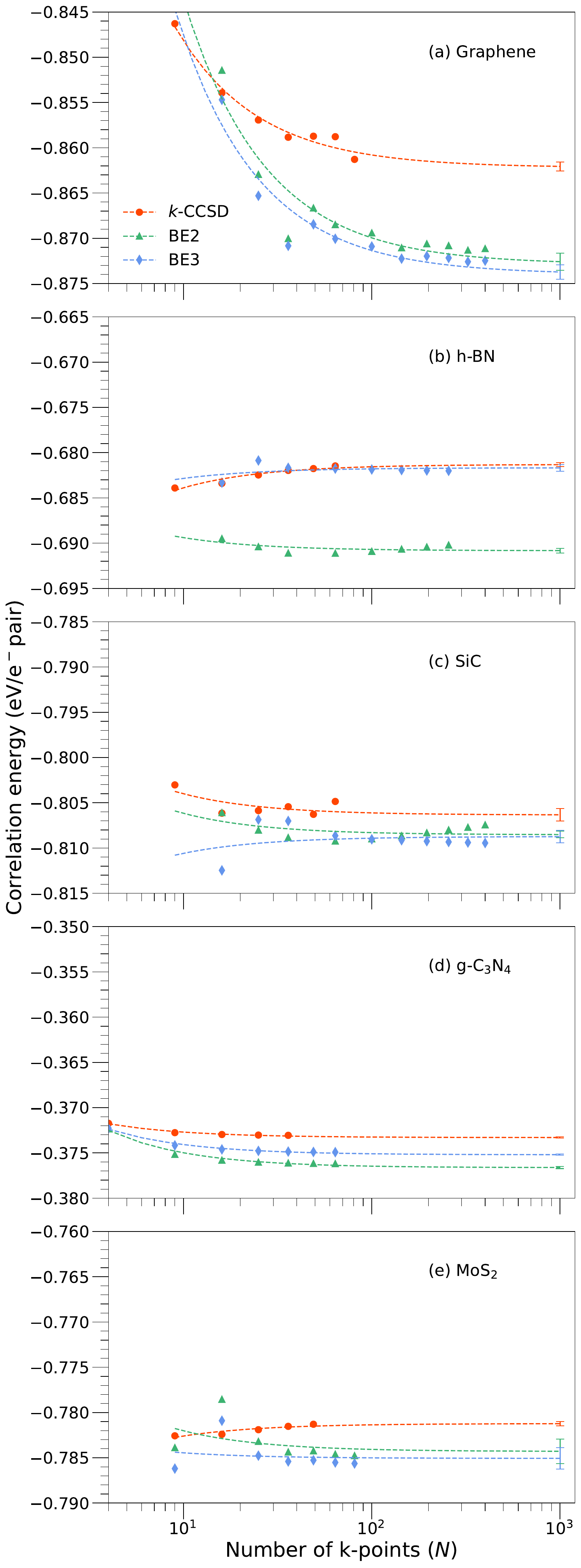}
\caption{Convergence of the total electron correlation energy to the thermodynamic limit with respect to $k$-points for 2D (a) Graphene, (b) h-BN, (c) SiC, (d) C$_3$N$_4$, and (e) MoS$_2$}\label{fig:conv_plot}
\end{figure}

\section{Results}
\subsection{Correlation Energies}
First, we apply BE-CCSD to calculate total correlation energies per unit cell for a diverse set of five 2D systems: graphene, h-BN, SiC, g-C$_3$N$_4$, and MoS$_2$. 
Graphene is a widely recognized zero bandgap semi-metal with unique electronic and thermal properties\cite{graphene1, graphene2}.
In contrast, h-BN, despite sharing the same honeycomb lattice structure with graphene, exhibits excellent insulating properties with a direct wide bandgap\cite{bn2}. Additionally, the other 2D systems namely SiC, g-C$_3$N$_4$, and MoS$_2$, emerge as direct medium bandgap semiconductorsy\cite{sic3, c3n41, mos22}. Apart from the electronic properties, these 2D systems exhibit high thermal and chemical stability, and are promising catalyst candidates\cite{graphene3, graphene4, bn1, c3n41}. Notably, g-C$_3$N$_4$, has shown significant effectiveness as a photo-catalysts across a broad variety of reactions. In terms of dielectric constants, graphene stands out with the samllest value of 2.2, while the other 2D systems exhibit dielectric constants ranging from approximately 6 to 15. The dielectric constant as well as the band gaps of the 2D systems are listed in Table \ref{tab:bandgap}.

\begin{table}[htb]
\caption{Error at the thermodynamic limit. The total electron correlation energy per unit-cell from the BE2 and BE3 are compared to the full k-point CCSD correlation energy. Absolute errors are in meV/e$^-$ pair}
    \begin{tabular}{l@{\hspace{1mm}}c@{\hspace{1mm}}c@{\hspace{3mm}}cc@{\hspace{5mm}}c@{\hspace{3mm}}c}
    \Xhline{2\arrayrulewidth}\\
            &&\multicolumn{2}{c}{BE2 error} && \multicolumn{2}{c}{BE3 error}\\[1mm]
            && Abs. & \%    &&  Abs.     & \%      \\[1mm]  
    \Xhline{2\arrayrulewidth}\\  
    Graphene     && 1.07$\pm$0.11 & -1.2$\pm$0.1 && 1.18$\pm$0.09  & -1.4$\pm$0.1\\[2mm]
    h-BN         && 0.95$\pm$0.03 & -1.4$\pm$0.1 && 0.04$\pm$0.04  & -0.1$\pm$0.1\\[2mm]
    SiC          && 0.22$\pm$0.08 & -0.3$\pm$0.1 && 0.24$\pm$0.10  & -0.3$\pm$0.1\\[2mm]
    g-C$_3$N$_4$ && 0.32$\pm$0.11 & -0.9$\pm$0.3 && 0.19$\pm$0.04  & -0.5$\pm$0.1\\[2mm]
    MoS$_2$      && 0.31$\pm$0.14 & -0.4$\pm$0.2 && 0.38$\pm$0.12  & -0.5$\pm$0.2\\[3mm]
    \Xhline{2\arrayrulewidth}
    \end{tabular}
    \label{tab:error}    
\end{table}

\begin{figure*}[hbt]
\includegraphics[scale=0.35]{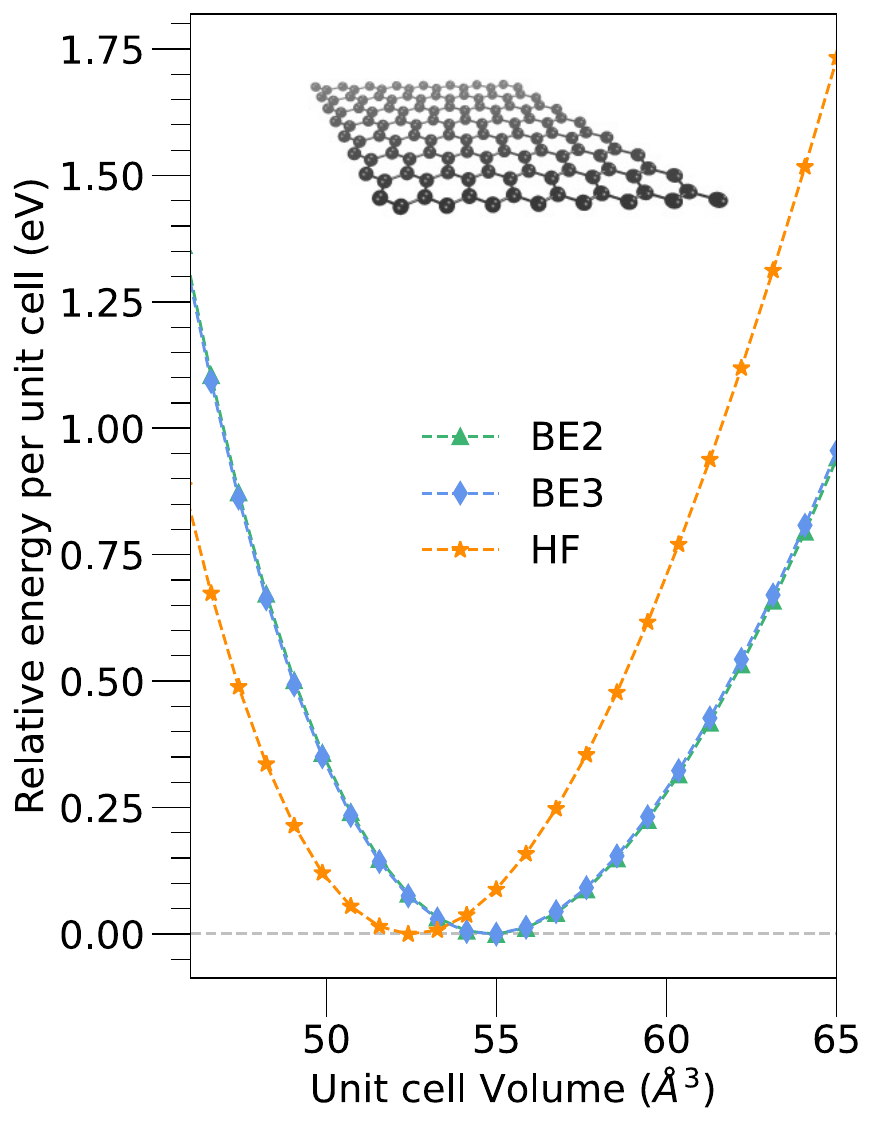}
\includegraphics[scale=0.35]{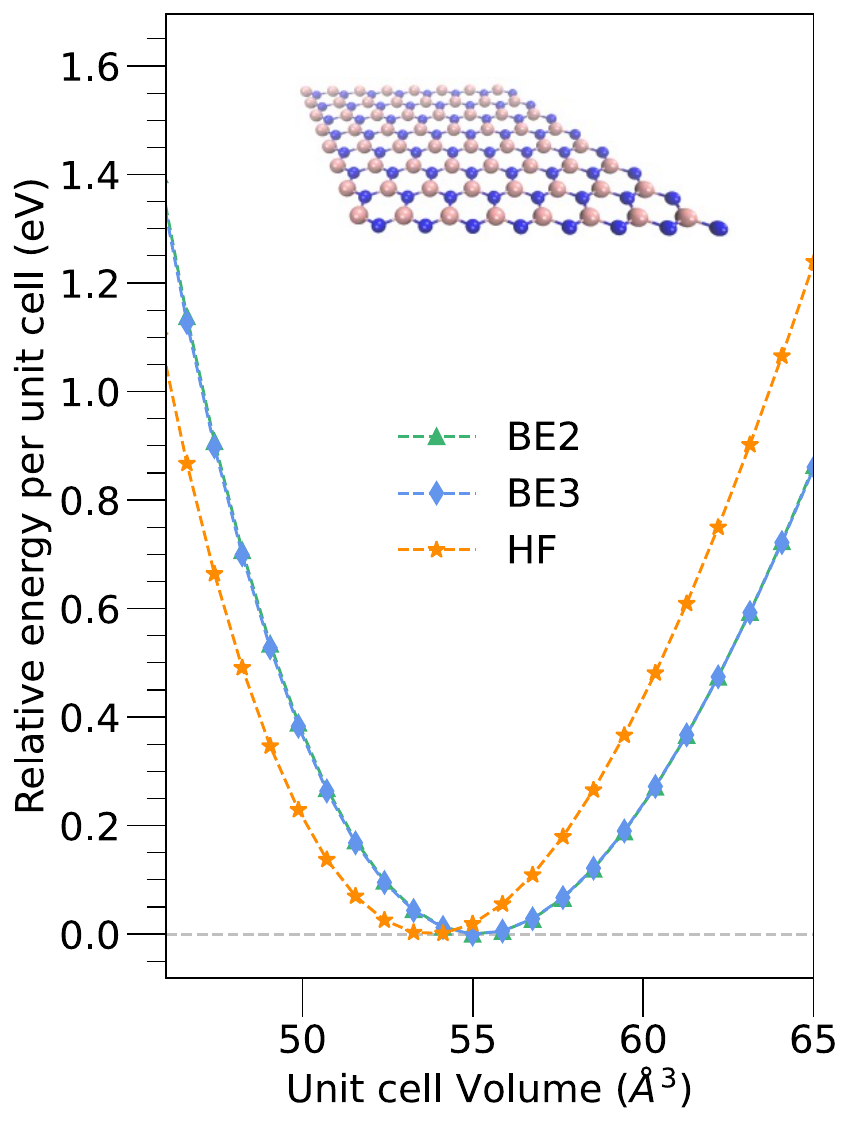}
\includegraphics[scale=0.35]{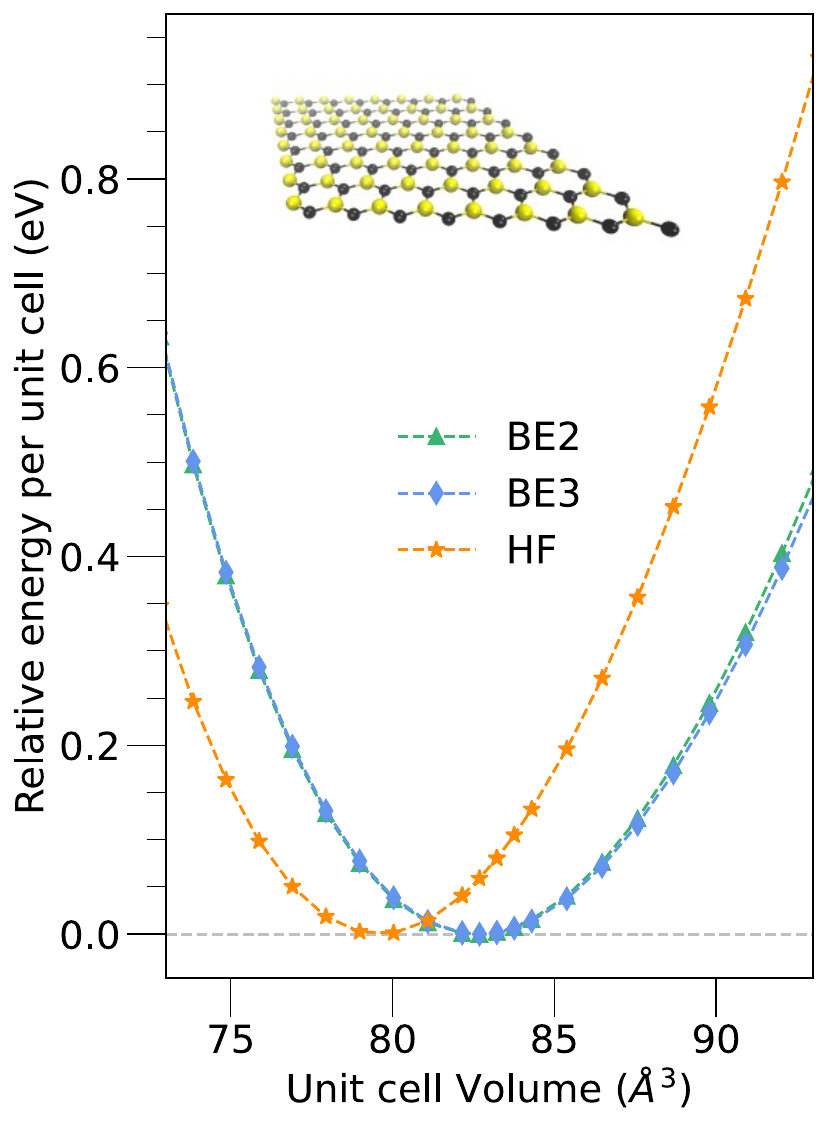}
\caption{Potential energy curves for graphene (left), h-BN (center), and SiC (right). The total energy per unit cell were computed using a $k$-mesh of (18$\times$18).}\label{fig:eos}
\end{figure*}

Table \ref{tab:error} presents the absolute and percentage error in the total electron correlation energy per unit cell at the TDL from BE2 and BE3 compared to the full $k$-CCSD energies. 
BE2 and BE3 reproduce the k-CCSD correlation with an error of -1.2$\pm$0.1\% and -1.4$\pm$0.1\%, respectively, for graphene. 
Conversely, for h-BN, BE3 yields accurate electron correlation energy with an error of only -0.1$\pm$0.1\% and BE2 with an error of -1.4$\pm$0.1\%. In the case of SiC, both BE2 and BE3 have an error of -0.3$\pm$0.1\%. For g-C$_3$N$_4$ and MoS$_2$, BE3 correlation energy has an error of -0.5\% with an uncertainty of $\pm$0.1 and $\pm$0.2, respectively, while BE2 results in an error of -0.9$\pm$0.3\% and -0.4$\pm$0.2\%, respectively. Overall, the BE3 correlation energies are observed to be better (or in some cases similar) than the BE2 correlation energies. BE3 consistently achieves an accuracy of $\sim$0.5\% (except for graphene), 
while the accuracy of BE2 shows less consistency, typically with an error of $\sim$1\%.

The accurary of BE3 correlates with the band-gap characteristics of the 2D systems. The corresponding band-gap values, along with the absolute errors of BE$n$, are provided in Table \ref{tab:bandgap}. Notably, BE3 almost exactly reproduce the full $k$-CCSD correlation energy for the h-BN, characterized by its wide band-gap. In contrast, the performance of BE3 is less accurate when applied to graphene which exhibits a zero band-gap. For 2D systems with intermediate band-gaps, specifically SiC, g-C$_3$N$_4$, and MoS$_2$, BE3 demonstrates a reasonable level of accuracy, falling between its performance on h-BN and graphene. Thus, BE3 yields accurate correlation energies for 2D systems characterized by medium to wide band-gaps, indicating that BE3 captures dynamic correlation more effectively compared to static correlation.

When considering dielectric constants of the 2D systems, there is no apparent correlation with the performance of BE$n$. It is worth noting that graphene, which has the lowest dielectric constant among the 2D systems studied in this work, demonstrates the least accurate results. Convergely, BE3 is most accurate when applied to h-BN, which exhibits a dielectric constant of 6.83. In comparison, BE3 when applied to  MoS$_2$, with a dielectic constant nearly twice that of h-BN, demonstrates an intermediate level of accuracy. The dielectric constant is associated with the range of correlation. However, given the lack of appratent dependence of the performance of BE$n$ on the dielectric constants and considering the sizes of the fragments for BE3, the range of correlation appears to have minimal influence on the performce of BE3.

\begin{table}[htb!]
\caption{Band-gaps, dielectric constant and absolute erros of BE$n$ correlation energies in meV/e$^-$ pair.}
    \begin{tabular}{l@{\hspace{1mm}}c@{\hspace{1mm}}c@{\hspace{3mm}}c@{\hspace{5mm}}c@{\hspace{7mm}}c@{\hspace{2mm}}}
      \Xhline{2\arrayrulewidth}\\
      && Band-gap & Dielectric & \multicolumn{2}{c}{Abs. error}\\[1mm]
      &&          & constant   & BE2 & BE3 \\[1mm]
      \Xhline{2\arrayrulewidth}\\  
      Graphene     && 0.0\cite{graphene1} & 2.20\cite{Elias2011} & 1.07 & 1.18\\[2mm]
      h-BN         && 5.9\cite{bn2} & 6.83\cite{Laturia2018} & 0.95 & 0.04\\[2mm]
      SiC          && 2.9\cite{sic3} & 9.70\cite{Liu_2022} & 0.22 & 0.24\\[2mm]
      g-C$_3$N$_4$ && 2.7\cite{c3n41} & 8.05\cite{10.1063/5.0045911} & 0.32 & 0.19\\[2mm]
      MoS$_2$      && 1.9\cite{mos22} &15.50\cite{Laturia2018} & 0.31 & 0.38\\[3mm]
\Xhline{2\arrayrulewidth}
    \end{tabular}
\label{tab:bandgap}
\end{table}

Figure \ref{fig:conv_plot} plots the convergence of the total correlation energy per unit cell with respect to k-points computed with BE$n$ and $k$-CCSD for the different 2D systems, re-scaled per electron pair. Despite employing the TA method for $k$-CCSD, obtaining a smooth correlation energy extrapolation to the TDL is challenging for graphene and SiC. In particular, the slower convergence of correlation energy observed in graphene, coupled with the non-smooth extraplolation to the TDL, introduces uncertainties when using the extrapolated TDL correlation energy as a reference to assess the accuracy of BE$n$. 
With more accurate $k$-CCSD correlation energy as a reference, it is plausible that these errors might be reduced. 
It is worth mentioning that correlation energy at extremely dense $k$-points, which are converged to the TDL can be computed with BE$n$, as depicted in Figure \ref{fig:conv_plot}. This eliminates the necessity for finite-size correction methods like TA in BE$n$ calculations.

\subsection{Lattice Constants and Bulk Moduli}
Having established the accuracy of periodic BE for correlation energies, we demonstrate applications of BE to obtain lattice constants and bulk moduli. Using a dense $k$-mesh of ($18\times18$) and minimal basis, we computed the total energy per unit cell at varying unit cell volumes of graphene, h-BN, and SiC. We then fit the results to a third-order Birch-Murnaghan equation-of-state\cite{eos_fit}, shown in Figure \ref{fig:eos}, to obtain the bulk modulus, $B_0$, and the lattice constant $a_0$. 
We note that minimal basis set calculations presented here demonstrate that periodic BE can be used to predict structural properties at highly dense $k$-points, where it approaches convergence to the TDL. 
Such computations would be very challenging with the full $k$-CCSD for these systems.

\begin{table}[htb]
\caption{Lattice constant $a_0$, and bulk modulus $B_0$ of graphene, h-BN, and SiC.}
    \begin{tabular}{l@{\hspace{4mm}}l@{\hspace{3mm}}cc@{\hspace{5mm}}l@{\hspace{3mm}}cc@{\hspace{5mm}}l@{\hspace{3mm}}c}
    \Xhline{2\arrayrulewidth}\\
            &\multicolumn{2}{c}{Graphene} & &\multicolumn{2}{c}{h-BN} && \multicolumn{2}{c}{SiC}\\[3mm]
            & $a_0$ & $B_0$ && $a_0$ & $B_0$&& $a_0$ & $B_0$ \\[1mm]
            & (\AA) & (GPa) && (\AA) & (GPa)&& (\AA) & (GPa) \\[3mm] 
    \Xhline{2\arrayrulewidth}\\  
        HF  & 2.46 & 265.54 && 2.49 & 231.02 && 3.03 & 167.57\\[2mm]
        BE2 & 2.52 & 215.36 && 2.52 & 207.57 && 3.09 & 143.07\\[2mm]
        BE3 & 2.52 & 216.30 && 2.52 & 206.24 && 3.09 & 140.51\\[2mm]
        Expt. & 2.46\cite{graphenel1} & && 2.5\cite{bnl} &&& 3.1\cite{sic1}\\[3mm]
    \Xhline{2\arrayrulewidth}
    \end{tabular}
    \label{tab:lattice_const}    
\end{table}

Table \ref{tab:lattice_const} presents the lattice constants and bulk moduli obtained from HF and correlated BE$n$ calculations. Although the difference between the lattice constant values obtained from HF and BE$n$ may seem small, it is essential to note that electron correlation impacts the structural properties of these systems. This is especially noticeable in the bulk moduli, where the electron correlation effect is more pronounced. The BE$n$ lattice constants are insensitive to the fragment size. BE2 and BE3 results only differ by less than a tenth of a picometer. However, a slight difference is observed in the BE2 and BE3 results for bulk moduli. In the case of graphene, the lattice constant obtained from HF reproduces the experimental value\cite{graphenel1}, which is slightly different from the BE$n$ lattice constant. 
However, the comparison to experiment is challenging due to the small basis set. Different conclusions entirely might hold in a more realistic basis set
We have not pursued this further as the results here are only proof-of-concept.
For h-BN, both HF and BE$n$ agree well with the experimental lattice constant\cite{bnl}. Finally, for SiC, the lattice constants obtained from BE$n$ differ by only a picometer from the experimental value\cite{sic1}. Overall, we observe that any errors associated with fragment size in BE$n$ are very smooth when comparing different geometries to one another - resulting in reliable and precise prediction of structural properties.

\begin{figure}[hbt]
\includegraphics[scale=0.35]{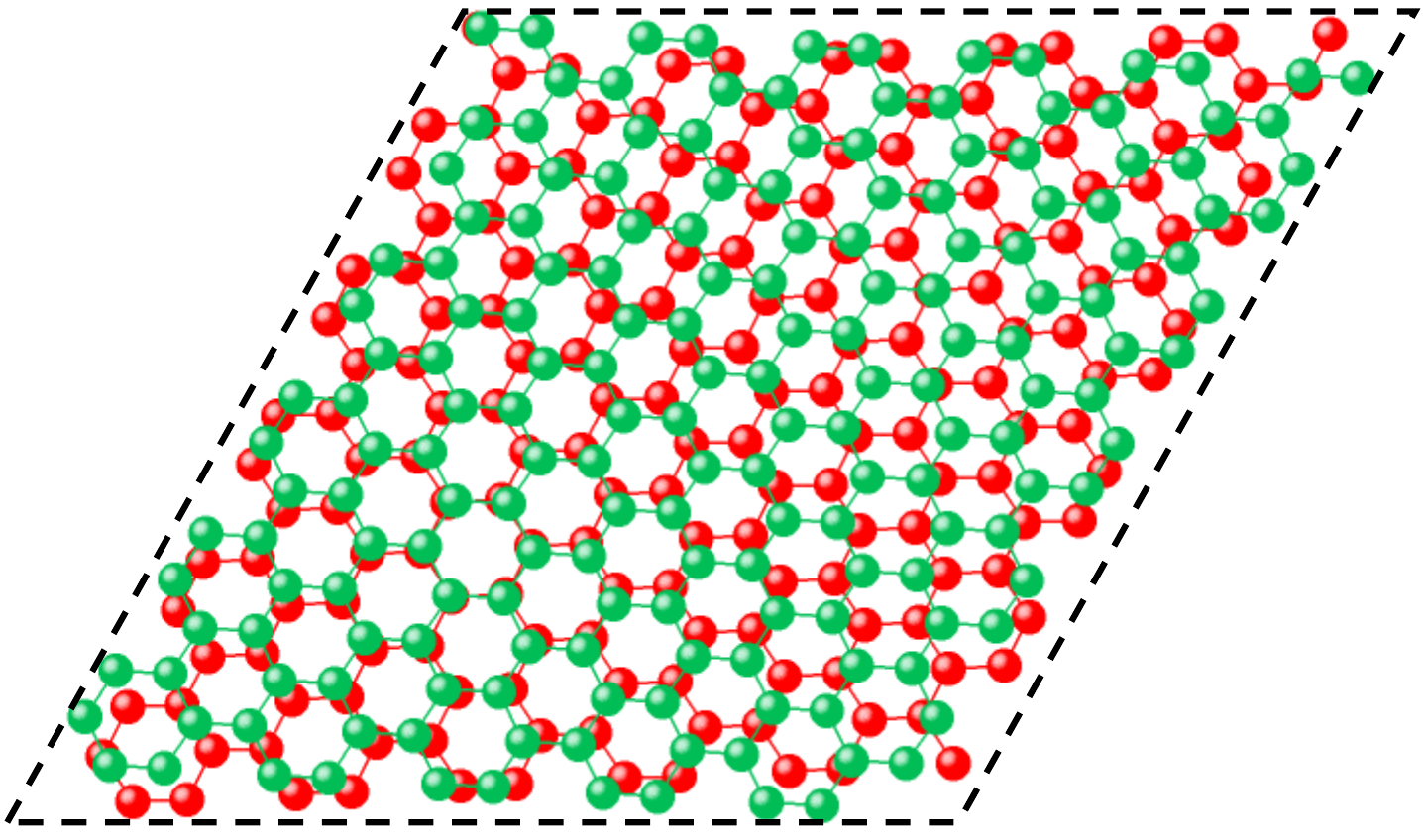}
\caption{Unit cell of twisted bilayer graphene superlattice with a twist angle of 6.01$^\circ$. The stacked monolayers of graphene are shown in red and black. The unit cell of the superlattice contains 364 carbon atoms.}\label{fig:tbg}
\end{figure}

\begin{figure}[hbt]
\includegraphics[scale=0.35]{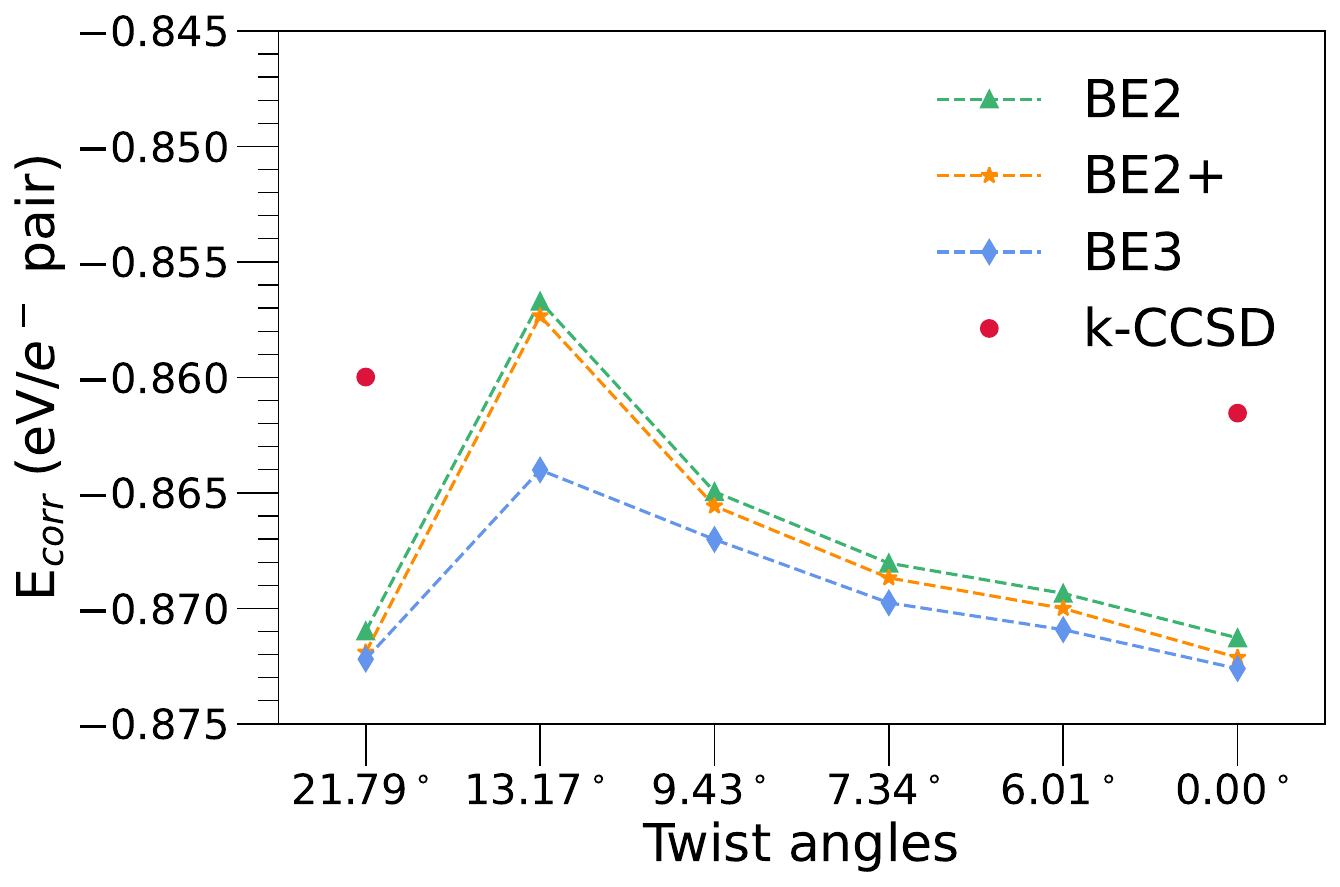}
\caption{Total correlation energy per unit cell computed with BE2, BE2+, and BE3 of the twisted bilayer graphene superlattice at different twist angles.}\label{fig:tbg_ecorr}
\end{figure}

\subsection{Twisted Bilayer Graphene}
Twisted bilayer graphene (TBG) has recently gained significant attention due to its extraordinary electronic properties resulting from a high periodicity moiré pattern\cite{bilayer_kaxiras, bilayer_kaxiras2, bilayer4}. This fascinating material is created by stacking two graphene monolayers with a specific twisting angle, resulting in a unique electronic structure that can be tuned by adjusting the twist angle\cite{Dean2019, bilayer_kaxiras}. The high periodicity moiré pattern observed in TBG leads to the emergence of new electronic states, including flat bands, which exhibit unique transport properties that make it a promising candidate for various electronic and optoelectronic applications\cite{Hyunmin2020, TIUTIUNNYK201936}. 
However, the honeycomb lattices of the two graphene monolayers only align to form periodic superlattices at certain discrete commensurate twist angles, and the elementary unit cell of the superlattices contains large number of carbon atoms\cite{MacDonald2011, bilayer_kaxiras}. 
In particular, near the “magic angle” of 1.5$^\circ$ where exotic electronic states are observed\cite{Dean2019}, the commensurate supercells contain tens of thousands of carbon atoms.
Even at relatively large twist angles, the number of atoms in the unit cell can be substantial. For instance, at approximately 6$^\circ$ twist angle, illustrated in Figure \ref{fig:tbg}, the unit cell contains 364 atoms. Computations with such large unit cells are impossible with the full k-CCSD and challenging even at the Hartree-Fock level.

\begin{figure}[htb]
\includegraphics[scale=0.7]{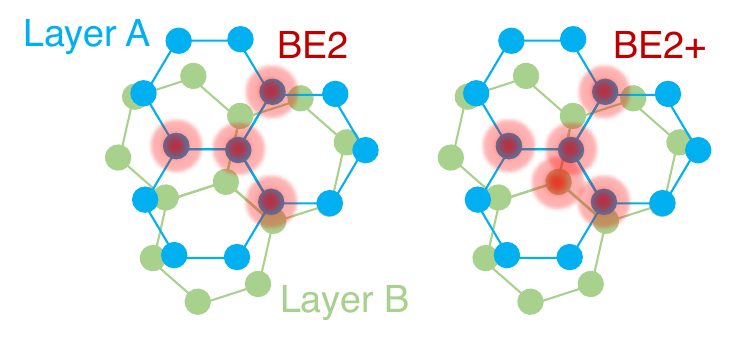}
\caption{Illustration of fragmentation schemes in bilayer graphene. BE$n$ fragment (left) comprises atoms from the same layer, while the BE2+ fragment (right) includes an additional atom from the adjacent layer.}\label{Fig:frag-bilayer}
\end{figure}

We utilized BE$n$ to compute the total electron correlation energy of TBG at twist angles of 21.79$^\circ$, 13.17$^\circ$, 9.43$^\circ$, 7.34$^\circ$, and 6.01$^\circ$. 
For twist angles smaller than 6$^\circ$, the basic HF calculations became prohibitive with the PySCF periodic HF code we used to perform our calculations. With a more efficient HF implementation, we estimate the BE2 and BE3 calculations would be feasible down to approximately 3.48$^\circ$, with a supercell size of 1084 atoms. Beyond that, the BE$n$ calculations would become disk-limited in performing the local integral transforms required for the fragment Hamiltonian. 
In addition to the BE$n$ fragment schemes, which consist of connected atoms, we introduce an additional fragment scheme called BE2+ (illustrated in Figure \ref{Fig:frag-bilayer}). This scheme incorporates an extra atom from the adjacent layer to capture the dispersion interactions between graphene layers. To achieve this, we employ a distance-based fragment scheme where we include the atom from the adjacent layer that is closest to the center of the BE2 fragment.

Figure \ref{fig:tbg_ecorr} presents the total BE correlation energy per unit cell at the different twist angles including the untwisted bilayer graphene, normalized per electron pair. We also performed $k$-CCSD computations for the untwisted bilayer graphene (twist angle of 0$^\circ$) and the twist angle of 21.79$^\circ$ employing the TA method. 
Because the $k$-CCSD computations were limited to coarse-grained $k$-points, we made comparisons for the BE$n$ energies computed at dense $k$-points to the $k$-CCSD energies at the coarse-grained $k$-points.
Specifically, for the twist angle of 21.79$^\circ$, the $k$-mesh was a 2$\times$2, while for the 0$^\circ$ twist angle, it was a 6$\times$6 $k$-mesh in the  $k$-CCSD calculations. For BE$n$, the $k$-mesh was 6$\times$6 and 18$\times$18 for the twist angle of 21.79$^\circ$ and 0$^\circ$, respectively.
For the 21.79$^\circ$(0$^\circ$) twist angle, BE2, BE2+, and BE3 exhibited errors of -1.3\%(-1.1\%), -1.4\%(-1.3\%), and -1.4\%(-1.2\%), respectively.
The comparison of BE correlation energies to $k$-CCSD energies is limited due to finite size errors at the coarse-grained $k$-points and the computational scaling of $k$-CCSD.
Nevertheless, we observe an unexpected physical phenomenon, where the correlation energy per electron pair initially decreases in magnitude before eventually increasing in the expected fashion as the twist angle decreases toward the magic angle (where strong correlation effects can appear). It isn’t entirely clear what leads to this effect, but it is interesting to note that it can only be accessed computationally using the tools presented here. k-CCSD is only computationally feasible for twist angles in the weak correlation regime where the correlation energy is decreasing with decreasing twist angle. Only with the improved BEn hierarchy can one extend closer to the size refime where new physics can emerge.

\section{Conclusion}
In conclusion, our study highlights the potential of periodic BE as a highly accurate electronic structure tool for studying 2D materials. Our BE3 fragment scheme with CCSD as the correlated fragment solver typically reproduces $\sim$99.5\% of the total electron correlation energy obtained from $k$-CCSD computations in 2D semimetal, insulator, and semiconductors.
Notably, the correlated calculations in periodic BE are performed without the need for reciprocal space $k$-point sampling, which streamlines the computational efficiency. 
Our results demonstrate that BE offers reliable and precise prediction of lattice constants and bulk moduli for 2D systems. Particularly noteworthy is the capability of BE to generate smooth potential energy curves with respect to unit cell volumes at densely sampled $k$-points for 2D systems. 
Moreover, we demonstrate that BE3 can efficiently perform correlated calculations on twisted bilayer graphene superlattices with large unit cell sizes, with the only limiting factor being the HF calculations.
The collective results presented herein demonstrate that periodic BE is a promising method with potential for future application in 2D materials.
The accurate correlation energies derived from BE offer the means to calculate band-gaps through the computation of chemical potentials, as outlined in Ref. \citenum{PhysRevA_bandgap} and \citenum{PhysRevLett.49.1691_perdew}, which will be the objective of near future work. 
To enable realistic computations beyond the proof-of-concept minimal basis calculations, future efforts will also focus on developing BE for larger basis sets. Another direction for future work is enhancing the quantitative accuracy of the method for systems with narrow or gap-less band gaps, to improve the overall reliability of periodic BE.

\section*{Supplementary Information}
Supplementary Information is provided for: (i) Distribution of correlation energy at the offsets in twist-averaging for $k$-CCSD; (ii) Geometric data including lattice constants and Cartesian coordinates.

\begin{acknowledgments}
This work was funded by a grant from NSF (NSF CHE-2154938).
\end{acknowledgments}

\bibliographystyle{achemso}
\bibliography{kbe.bib}

\end{document}